# Discrete-Event Simulation in Healthcare Settings: a Review

John J. Forbus
Dept. of Systems Engineering
University of Arkansas at Little Rock
2801 S University Ave
Little Rock AR 72204
jjforbus@ualr.edu

Daniel Berleant
Dept. of Information Science
University of Arkansas at Little Rock
2801 S University Ave
Little Rock AR 72204
jjforbus@ualr.edu

**Abstract**

Background: We review and define the current state of the art as relating to discrete event simulation in healthcare-related systems.

Methods: A systematic review of published literature over the past five years (2017 - 2021) was conducted, building upon previously published work.  PubMed and EBSCOhost were searched for journal articles on discrete event simulation in healthcare resulting in identification of 933 unique articles. Of these about half were excluded at the title/abstract level and 154 at the full text level, leaving 311 papers to analyze. These were categorized, then analyzed by category and collectively to identify publication volume over time, disease focus, activity levels by country, software systems used, and sizes of healthcare unit under study.

Results: A total of 1196 articles were initially identified.  This list was narrowed down to 311 for systematic review.  Following the schema from prior systematic reviews, the articles fell into four broad categories: health care systems operations (HCSO), disease progression modeling (DPM), screening modeling (SM), and health behavior modeling (HBM).

Conclusions: We found that discrete event simulation in healthcare has continued to increase year-over-year, as well as expand into diverse areas of the healthcare system.  In addition, this study adds extra bibliometric dimensions to gain more insight into the details and nuances of how and where simulation is being used in healthcare.

Keywords:

Discrete event simulation, Healthcare applications, Systematic review

# 1. Background

## 1.1 Introduction

Simulation is "the imitation of the operation of a real-world process or system over time" (Banks, 1998). It is a simplified model of a system of interest, used to study specific aspects of the system. Simulations are used because they are so often faster, easier, less expensive, and less risky than real-world observations and experiments, allowing decision makers to test a variety of scenarios, make predictions, and study alternatives. Discrete event simulations are stochastic, dynamic, and discretely-changing (the system state changes at discrete points in time, as opposed to a continuously-changing state like those seen in system dynamics models).

Discrete event simulation (DES) has been used since the 1950s in diverse fields such as manufacturing, supply chain management, military operations, computer and network design, and even voting systems (Allen, 2011). As computer processing power expands, the resources necessary to model more complex systems becomes within the reach of an ever-increasing pool of potential applications, further highlighting its utility and power (Robinson, 2005).

DES, because it can model process flows that involve stochastic timings and decision trees (Crane, 1975), is ideally suited to a manufacturing environment, where assembly lines, product routings, and processing and queuing times are clearly defined (Babulak, 2010), and the variation in times at each step is well understood (Schroer, 1988). The healthcare industry is not wholly dissimilar from a manufacturing environment (Marcinko, 2014).

In typical manufacturing DES scenarios, raw materials enter the system, are routed to wait in queues, transformed and/or consumed via various processes, and finished goods exit the system. Similarly, in a healthcare setting, sick or injured patients enter the system, wait in waiting rooms, are transformed via various processes (surgery, radiological exam, course of treatment, etc.), and healthy patients exit the system. More recent advances in DES consider long-term population health outcomes of different treatments, therapies, or medications; the economic implications for quality-adjusted life years (QALY) are frequent subjects of study (Nguyen, Megiddo, & Howick, 2019).

## 1.2 The Need for DES in Healthcare

The healthcare industry, especially in the United States, is under increasing pressure to do more with less. In the US, hospitals and clinics find themselves forced away from a fee-for-service model and toward a value-based model, wherein the healthcare provider receives a set annual payment from various payers, such as Medicare/Medicaid and the state's private insurers, and then expected to maintain a healthy patient population (Lockner, 2018). While value-based care should ultimately lead to lower costs for financial consumers, it also means lower reimbursements for healthcare providers. This puts financial pressure on healthcare providers to increase volume and throughput without sacrificing patient health, staff safety, or population health outcomes (Dall, 2013).

These pressures are causing more healthcare providers to apply efficiency and quality improvement techniques to their workflows. Some of the more commonly-used tools are well-known in manufacturing engineering, such as Lean and Six Sigma (Koning, 2006). Both are well-suited to small- to medium-scale improvement projects with well-defined scopes and clear objectives. The use of the Deming Plan-Do-Study-Act improvement cycle is used extensively in healthcare improvement projects (Taylor, 2013). The PDSA cycle is an iterative improvement approach, wherein small proposed tests of change are tried in practice, the results studied, and revisions or expansions made based on the results (Donnelley, 2015). While potentially very powerful, it has the disadvantage of being prohibitively disruptive for large-scale changes. As more healthcare providers encounter this limitation, DES is being applied to a greater number of hypothetical situations.

The "what if" ability of DES to test many configurations of different possible changes allows decision makers to hone in on the leading candidates for practical application tests of change, without burdening the healthcare staff with non-value-added churn (Zhao, 2015). Yet, the healthcare industry is often considered particularly challenging for such quantitative analyses (Zhang, 2018). Against this reluctance to apply such a powerful tool, much work remains to demonstrate the utility of discrete event simulation to help build the healthcare systems of tomorrow.

Discrete event simulation is well suited to addressing a wide range of scenarios. One part of this range holds DES systems for optimizing resources, such as number of nurses or GPs needed by the unit. Another consists of DES systems for comparing the resource needs of alternative interventions, an approach in use particularly in the UK, Netherlands, Australia and Canada. However, a designer should understand both the advantages and disadvantages of DES in order to properly determine if it is the appropriate tool for the job.

**1.3 Advantages**

The ability of DES to try various scenarios in a virtual setting is far less expensive than having to commit resources to test in the real world (Banks, 1998). Additionally, a designer can test an almost limitless number of different scenarios, increasing the likelihood of finding the ideal configuration. Some software packages even have integrated linear programming tools, allowing for the discovery of true optima.

As part of the ability to model different scenarios, the designer can also use the tool to better understand a current system. The ability to either view years of operation in moments, or slow the system to a crawl, allows the designer to see into the inner workings of the system, to understand emergent system-level interactions, and to find event-level bottlenecks. This allows the designer to identify constraints, excess capacity, and uneven utilization, which can then be designed out for an improved system.

Specific to the study of healthcare systems, discrete event simulation has several advantages over other commonly-used analytical tools. A significant area of focus in healthcare DES is "health technology application," which is the long-term economic impact of various treatments on population health. For many years, Markov models were the tool of choice for such studies (Caro and Möller, 2016). While Markov models are computationally tractable, they are poorly suited to modeling patient populations with comorbidities. In contrast, discrete event simulation can model a patient population with a wide variety of symptoms and disease processes, which can change over time. Discrete event simulation models are also more accurate with respect to event timings, as they do not rely on defined system cycle time like Markov models, thereby producing more accurate results (Standfield, Comans, and Scuffham, 2015).

**1.4 Disadvantages**

Discrete event simulation has disadvantages as well. Model building requires specialized skills in coding and mathematics. Many dedicated software packages exist that ease the burden of coding, but they can incur a significant up-front cost, and each still has its own syntax and grammar which requires time and experience to master.

Data collection, modeling, and results analysis can each be time consuming and costly. Large simulations with multiple input variables can take many hours to resolve. A simulation model output is only as good as the data input that drives it, and without taking the time to ensure quality input data, the output may not yield usable results. Even with good input data, the results may still be difficult to interpret, especially for someone not accustomed to probabilistic outputs. Even for a well-trained designer, it can sometimes be difficult to distinguish if variation in outputs across runs is caused by the proposed system changes being studied, or just statistical noise (Sharma, 2015).

The purpose of this systematic literature review is to define the current state of the art with respect to the use and application of discrete-event simulation as applied to health care settings and delivery. As the pace of DES use increases (Zhang, 2018), so too should surveys of the literature, to stay abreast of what is being done, how, and importantly, potential areas for expansion. Additionally, we seek to expand upon prior review work by adding extra dimensions in this systematic review, in an attempt to determine if there are areas of study being overlooked by current research.

## 2. Methods

### 2.1 Literature search strategy

A literature search was conducted in the PubMed and EBSCOhost databases. The searches were conducted on 08 January 2022, and covered articles published in calendar years 2017 through 2021. This builds on the prior work by Zhang (2018), who compiled a systematic review of the literature through 31 March 2017. The search terms used were "discrete event," and "healthcare." The search was limited to published journal articles only; conference proceedings, books, magazines, and non-peer-reviewed papers were omitted. There were no limitations on language or country of origin; however, if an English translation was not available, the result was excluded.

To be included in the review, the articles had to meet the following criteria:
1) have healthcare delivery as the primary system of interest
2) use discrete-event simulation as the primary modeling tool
3) demonstrate a clearly-defined workflow of the system simulated as per the ISPOR-SMDM modeling good research practice (Marshall et al., 2015a, 2015b))
4) quantify the inputs to and results from the study
5) appear in peer-reviewed academic journals

### 2.2 Search Results

The PubMed search returned 467 results, while the EBSCOhost search returned 719 results. Of the initial pool of 1,186 articles, 253 results were duplicates. Another 622 were excluded at either the title-and-abstract level, or after the full text review: 351 because they were not focused on discrete event simulation; 120 because they were not focused on healthcare applications; 127 because they were meta-studies (such as literature reviews); 20 were not journal articles; four were not available in English. This left a total of 311 articles for the systematic review. This screening per the Preferred Reporting Items for Systematic Reviews and MetaAnalyses (PRISMA) guidelines. The PRISMA flow diagram is shown in Figure 1.

This systematic review is focused on the application of discrete event simulation to solving healthcare-related problems. As such, only limited information was retrieved from each article for the purpose of this analysis. Author, title, journal, date, simulation application, simulation focus, disease process, country, healthcare setting, facility (or study) size, subject age, and software used were the data points gleaned from each.

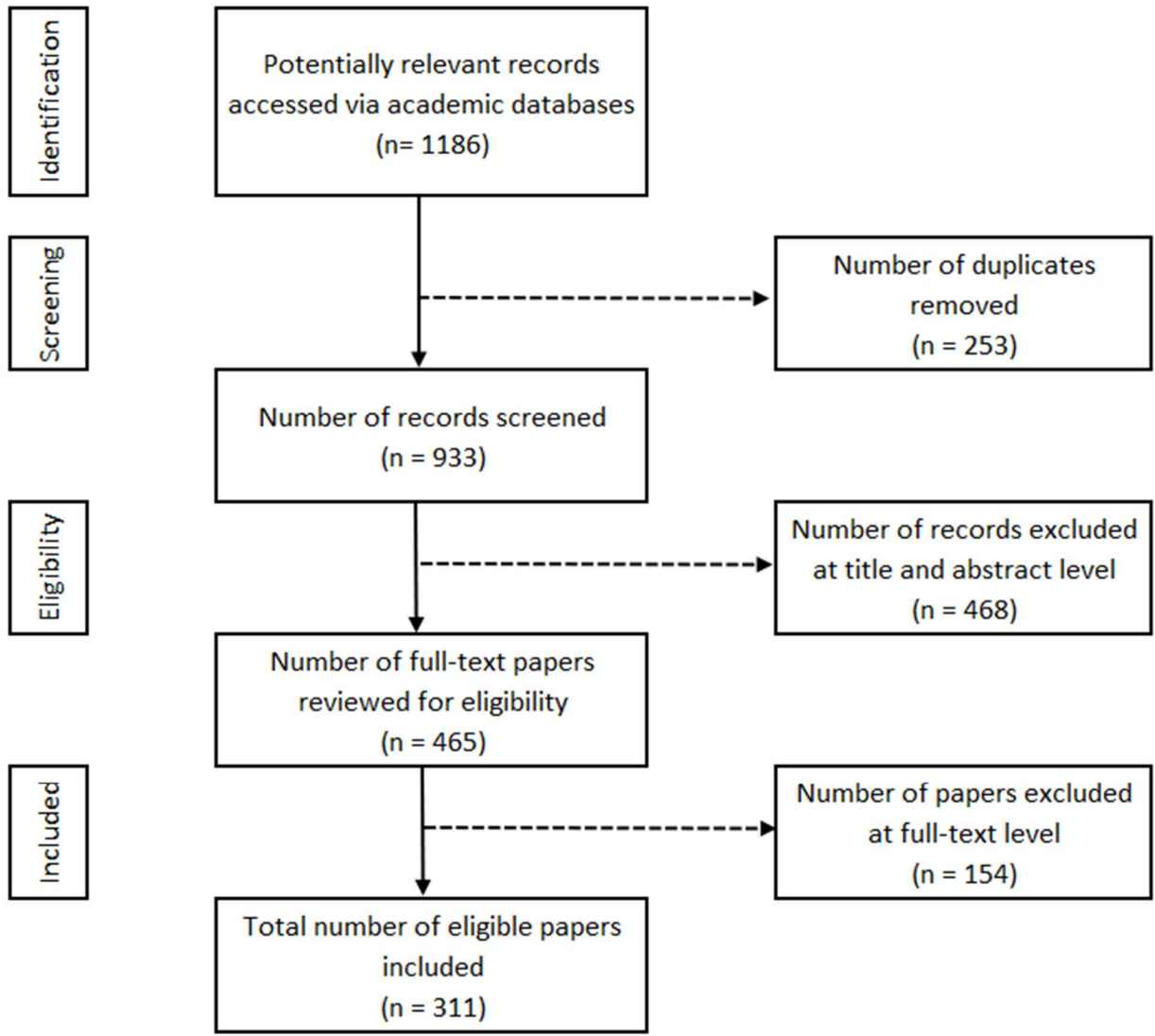

Figure 1. PRISMA flow chart of the systematic literature review

## 3. Results - Classification and analysis

The classification schema follows that outlined by Zhang (2018). It also expands upon that work by adding extra dimensions of study - software, location, size, and age. The simulation articles can be divided based on four primary purposes. The first is healthcare system operations (HCSO), akin to traditional manufacturing operations management. This is concerned with things such as resource utilization, scheduling, capacity planning, and system diagnosis. The remaining three address different kinds of healthcare intervention, but share a similar purpose from a health economics standpoint, namely, comparing cost and effectiveness of different healthcare deliveries without specifically addressing budgets or other resource. The second is disease progression management (DPM). This studies population health outcomes, looking at long-term economic effects of treatments and the relative merits of different care pathways. The third is health screening protocols, and the effects of properly steering populations into different care pathways. The final purpose is health behavior modeling (HBM), which studies diseases either caused or exacerbated by personal lifestyle choices, and how those choices can affect population health outcomes.

### 3.1 Healthcare DES over time

Over the time period of this review, the number of DES articles about healthcare applications (HCSO category) mostly rose each year. See Figure 2. For prior years the number of healthcare-focused DES studies also generally increased year-over-year as noted in Zhang's (2018) earlier review (Figure 3). A notable exception to the trend is 2021, which may have been impacted by the global pandemic, like much of the world economy. It will be interesting to see if that number resumes its climb in 2022 and beyond.

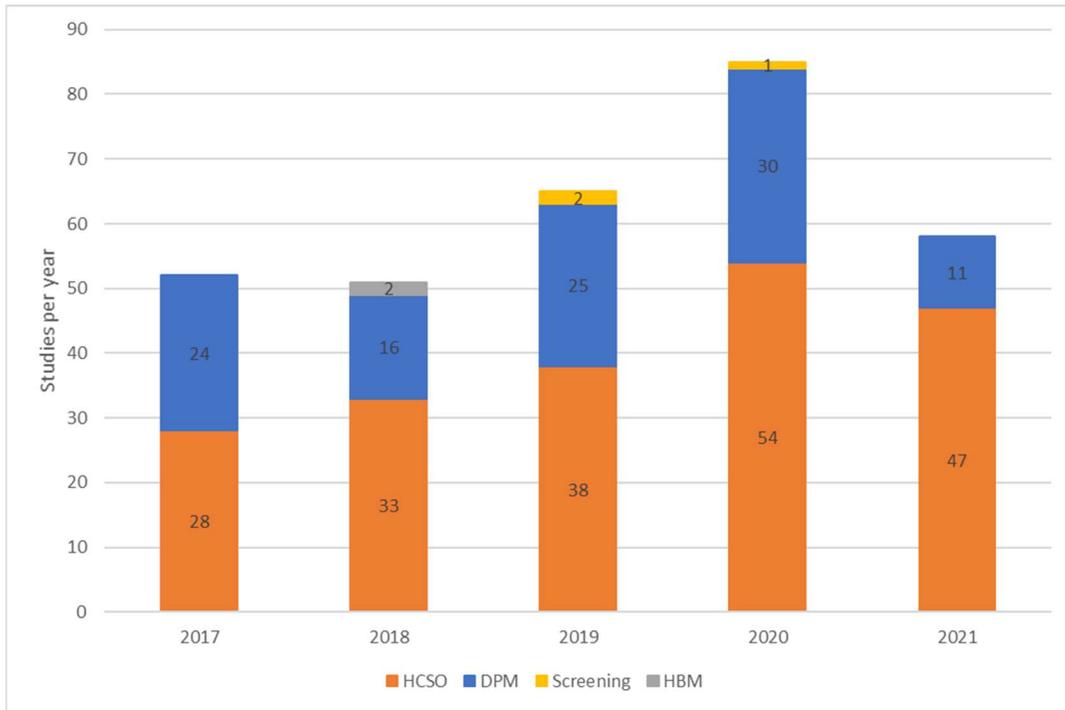

Figure 2. Healthcare DES articles per year and application

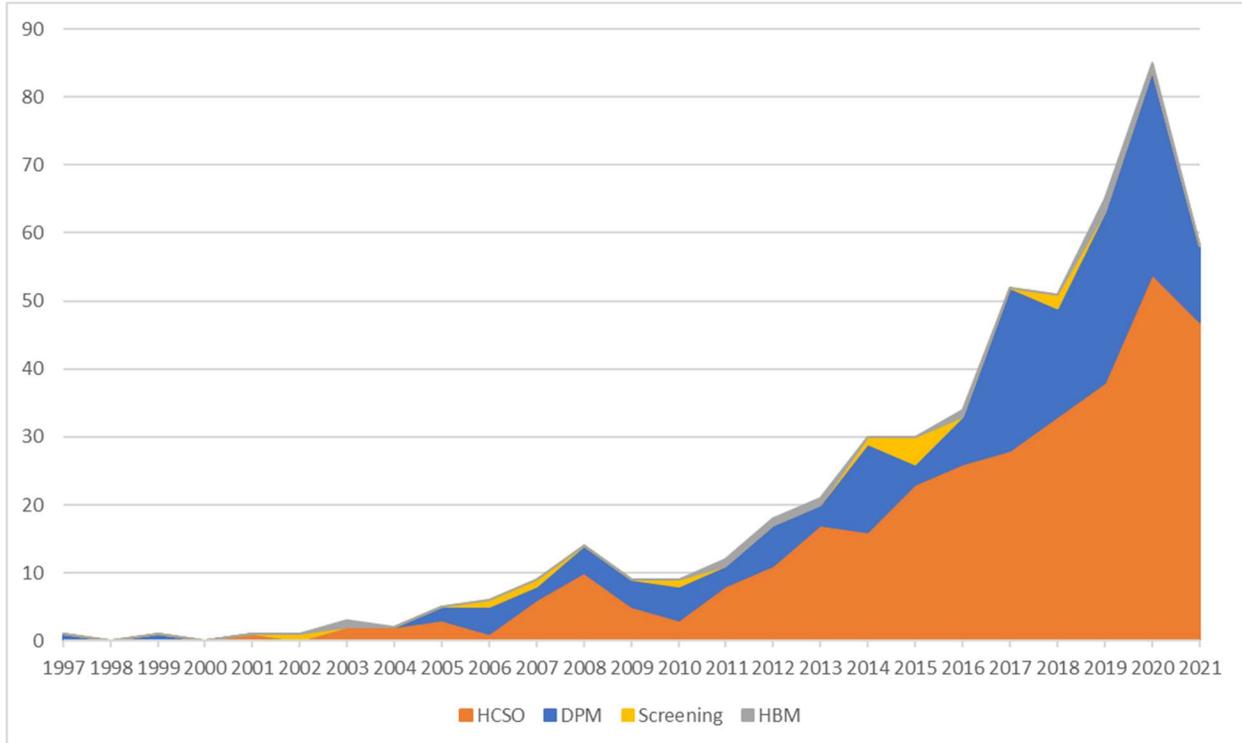

Figure 3. Healthcare DES articles over time

## 3.2 Studies by disease process

As mentioned previously, DES is typically employed to solve a specific problem. In healthcare, that is frequently either a specific disease process or specific functional area. Healthcare system operations (HCSO) tends to focus more on the functional area, while disease progression modeling (DPM), health behavioral modeling (HBM), and screening modeling focus on the disease process. Even in the case of HCSO, the disease process can factor into the type of study performed. Figure 4 shows the distribution of disease processes studied excluding HCSO articles; Figure 5 shows the total distribution with HCSO included.

.

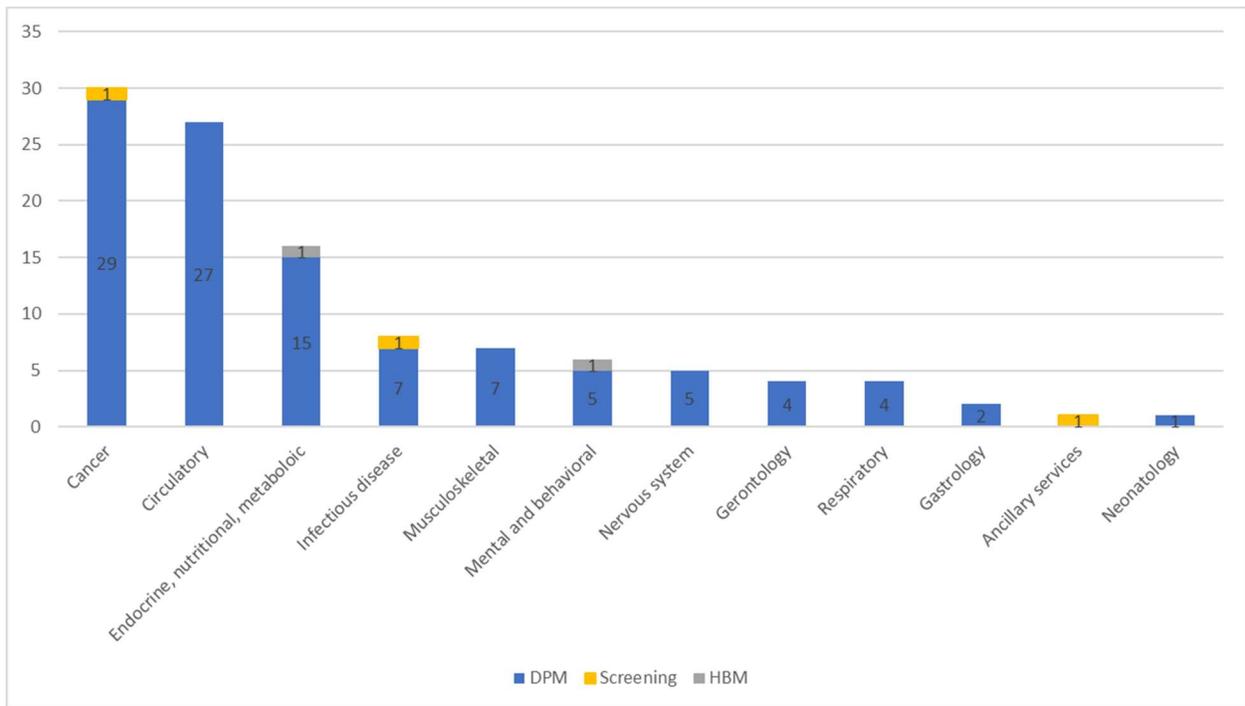

Figure 4. Studies by disease process (HCSO excluded)

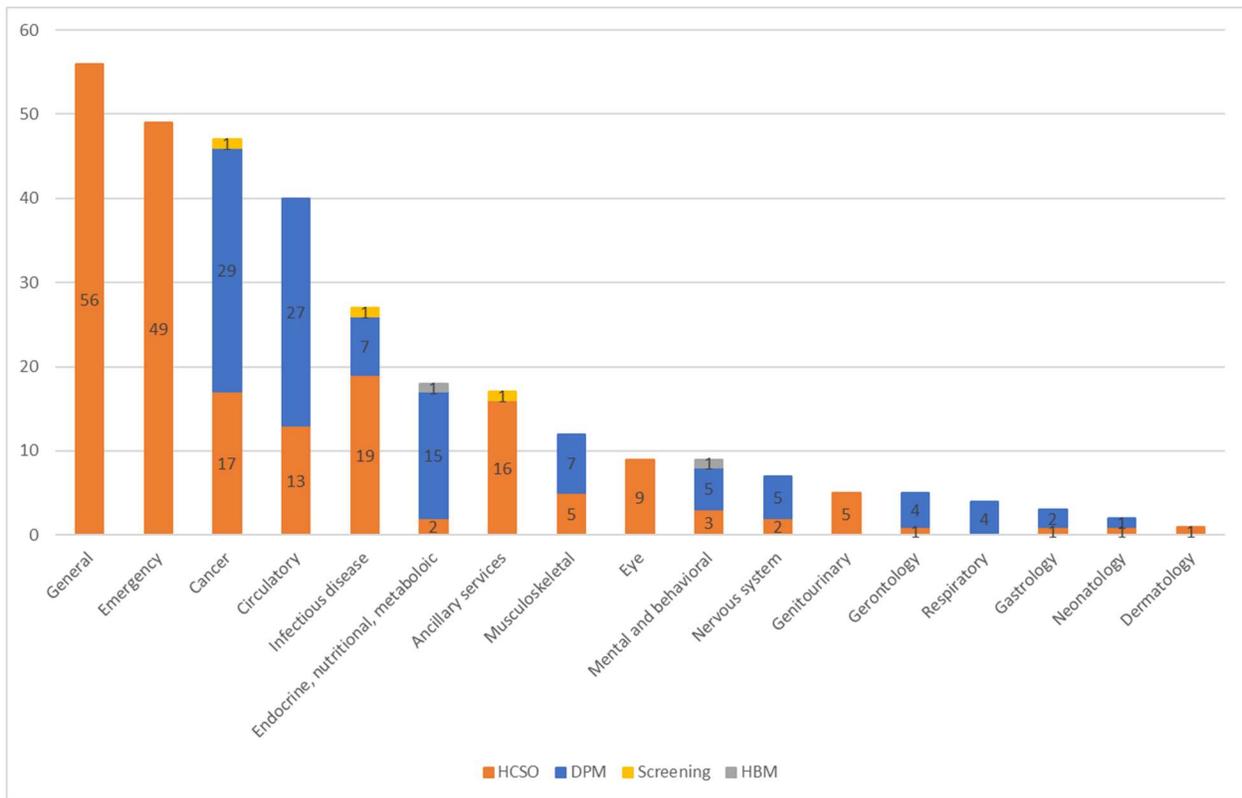

Figure 5. Studies by disease process (HCSO included)

Cancer (all types) and circulatory disease (including heart disease, heart attack, stroke, and thrombosis) are the most frequently studied disease progression models, perhaps because they affect large numbers of people each year.  Adding HCSO studies to the analysis, general medicine and emergency medicine make strong showings.  This can be explained by what most HCSO models study, which are hospital emergency departments, inpatient hospitals, and outpatient clinics, especially their flow and process optimization problems.  When viewed year-over-year, infectious disease models saw a large up-tick in 2020 and 2021, as more simulations were devoted to the COVID-19 pandemic, either in modeling the outbreak (DPM) (Bardet *et al*, 2021) or in capacity planning for surges or mass-vaccination clinics (HCSO) (Wood *et al*, 2021).

**3.3 Disease progression modeling**

When applied to disease progression modeling, DES is used to help understand the course of a disease on a large population over long periods of time.  The economic effects of treatments, typically including quality-adjusted life years (QALYs), are modeled so that researchers can make well-informed decisions about how to allocate treatment resources, and which treatments are most effective for the investment of time, effort, and capital.  There are four main areas of focus within DPM.  The most common, with 67 of 106 studies (63.2%), studies the long-term economic outcome and impact of specific treatments.  The second, with 21 of 106 studies (19.8%), contains treatment comparisons.  These look at the long-term prognosis of one care pathway versus another.  The third type, with 9 of 106 studies (8.5%), is screening models.  These involve identifying target populations for treatment and the outcomes of timely (or delayed) treatment.  The fourth type, with 8 of 106 studies (7.5%), projects demand for treatments and therapies over time.  One final study was unusual for DPM studies.  It was a system diagnosis study, comparing the effectiveness of different DES models at modeling disease progression (0.9%) (Willis *et al*, 2020).

Of those studies of disease progression modeling, the vast majority (100 of 106, 94.3%) were analyses at the population health level - that is, the total economic impact a health system could expect to encounter.  Two of 106 studies focused on multi-facility hospital systems (Dutta *et al*, 2020; Larh *et al*, 2020), and two on individual outpatient clinics (Bae *et al*, 2020; Ambavane *et al*, 2020) (1.8% each).  One of the 46 studies was focused on an inpatient hospital (Dieleman *et al*, 2020), and one on an emergency department (Soorapanth & Young, 2019) (0.9% each).

**3.4 Healthcare system operation**

Healthcare system operations (HCSO) attempts to help decision makers allocate constrained resources to meet demand in a timely and efficient manner by understanding process flow, system bottlenecks, and resource allocations.  Investigators can test different process configurations against each other, or evaluate different patient arrival rates or staffing patterns.

The functional area studied is more relevant to a discussion of HCSO than the disease process (Table 1).  It is beneficial to see what areas are being studied to identify demand.  Outpatient clinics were the largest plurality of studies, with 63 of 200 (31.5%).  The most-studied types of clinic were general/primary care and oncology, primarily concerned with resource allocation to reduce visit length and increase throughput.  Inpatient hospital operations were the next largest segment, with 62 of 200 studies (31.0%).  The individual departments studied varied, but the largest sub-percentage were multi-department studies, typically looking at total operational efficiency as a means to reduce inpatient length of stay.  The emergency department, with 49 of 200 studies (24.5%) was the third largest area of focus.  Emergency departments are often attached to inpatient hospitals, but the volume of ED studies was large enough to warrant its own category.  ED studies focused most often on resource allocation and effects of operational change, to trial different scenarios in simulations.  Capacity planning also featured prominently in ED studies, with many focused on surges during disasters.  Multi-facility hospital systems were modeled in 12 of 200 studies (6.0%), focusing on resource allocation and capacity planning.  The health care system as a whole was studied 11 times (5.5%), primarily for system diagnosis purposes (validating different DES models and techniques).  There was one study each of emergency medical transport (EMT) (Nawaz *et al*, 2018), a pharmacy (Furushima *et al*, 2018), and a medical supply distributor (Al-Fandi *et al*, 2019)  (0.5% each).

| Outpatient clinic | | 63 | 31.50% |
|---|---|---|---|
| | General | 12 | 6.00% |
| | Hematology Oncology | 8 | 4.00% |
| | Ophthalmology/Optometry | 8 | 4.00% |
| | Primary care | 6 | 3.00% |
| | Vaccine | 4 | 2.00% |
| | Ob/Gyn | 3 | 1.50% |
| | Vascular | 3 | 1.50% |
| | Ambulatory surgery | 2 | 1.00% |
| | Cardiac | 2 | 1.00% |
| | Orthopedics | 2 | 1.00% |
| | Telemedicine | 2 | 1.00% |
| | Urgent care | 2 | 1.00% |
| | Laboratory | 2 | 1.00% |
| | Dematology | 1 | 0.50% |
| | Dentistry | 1 | 0.50% |
| | Maternal/child health | 1 | 0.50% |
| | Neurology | 1 | 0.50% |
| | Pediatric | 1 | 0.50% |
| | Radiology | 1 | 0.50% |
| | Sexual health | 1 | 0.50% |
| Emergency Department | | 49 | 24.50% |
| Multi-facility provider | | 12 | 6.00% |
| Healthcare System | | 11 | 5.50% |

| Inpatient hospital | | 62 | 31.00% |
|---|---|---|---|
| | Multi-department | 17 | 8.50% |
| | Surgery | 13 | 6.50% |
| | MedSurg unit | 6 | 3.00% |
| | ICU | 5 | 2.50% |
| | Radiology | 5 | 2.50% |
| | General | 4 | 2.00% |
| | Acute care | 1 | 0.50% |
| | Adminstration | 1 | 0.50% |
| | Cardiac | 1 | 0.50% |
| | Gerentology | 1 | 0.50% |
| | Hospice care | 1 | 0.50% |
| | NICU | 1 | 0.50% |
| | Oncology | 1 | 0.50% |
| | Pathology lab | 1 | 0.50% |
| | Pediatric | 1 | 0.50% |
| | Pharmacy | 1 | 0.50% |
| | Psychiatric ward | 1 | 0.50% |
| | Dialysis | 1 | 0.50% |
| Pharmacy (Stand-alone) | | 1 | 0.50% |
| Emergency Medical Service | | 1 | 0.50% |
| Distributor (Medical Warehouse) | | 1 | 0.50% |

Table 1. HCSO functional areas

### 3.4.1 HCSO sub-categories

Within the set of HCSO studies, it is informative to study the focus areas (Figure 6). Of the 200 HCSO studies, 66 (33.0%) were focused on resource allocation. Those dealt with topics such as maximizing throughput, minimizing wait time, and improving operational efficiency. Forty-four of the studies (22.0%) dealt with the effects of operational change, comparing two or more competing solutions against one another to determine which (if either) should be implemented. Forty of the studies (20.0%) were capacity planning studies, evaluating the system's ability to handle possible changes in volume. Nineteen of the studies (9.5%) evaluated different patient arrival templates, and thirteen (6.5%) evaluated different staffing patterns. Thirteen (6.5%) were system diagnosis evaluations. These compared the results of discrete event simulation models to real-world data to validate which of the models is better suited to use in future studies. Only five (2.5%) were purely economic analyses. Thus resource allocation, capacity planning, and effects of change were the most commonly-used applications within HCSO simulations. Economics and scheduling both lagged behind.

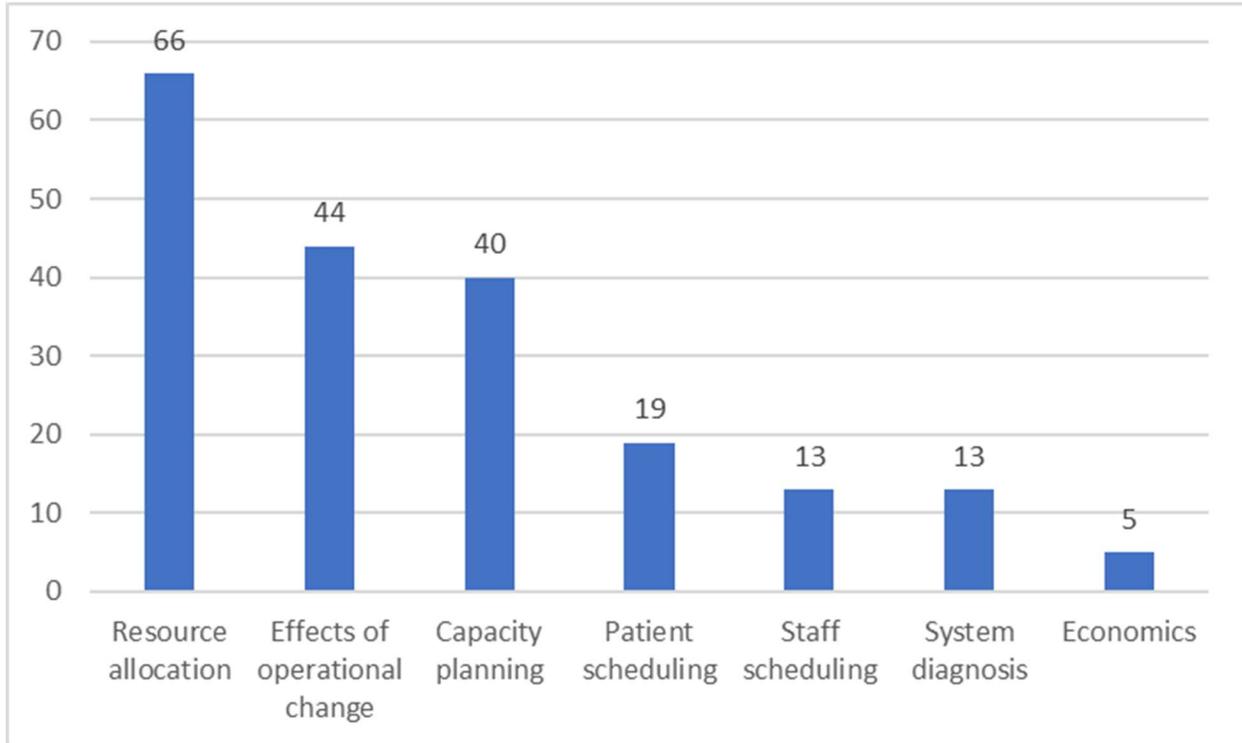

Figure 6. HSCO areas of focus

### 3.5 Screening modeling

Three studies were identified as discrete event simulations of screening models. These studies looked at the potential effects of changing the methods by which someone is identified as having a particular disease or qualifying for a particular treatment pathway. One studied cancer screening (Sewell *et al*, 2020); a second tuberculosis (Campbell *et al*, 2019); the third genotyping (Shi *et al*, 2019). Each compared the long-term economic impact of screening potential patients for care, and the resultant QALYs that treatment might generate.

### 3.6 Human behavior modeling

Human behavior modeling is a relatively new field in discrete event simulation. At an individual level, human agents can be difficult to model. They have intrinsic motivators and may make seemingly-irrational decisions, which can be difficult to model. Only two of the studies were found to deal with human behavior modeling. One looked at policies that might affect rates of obesity (Nau *et al*, 2018); the other compared smoking cessation strategies (Pennington, Filby, Owen, & Taylor, 2018). Both looked at economic impacts of the proposed interventions and the resultant QALYs.

### 3.7 Location distribution

The locations of the health systems studies track generally with the population and level of technological sophistication of the country. See Figure 7. It should not be surprising that the United States and United Kingdom have the most published studies, with totals of 72 and 66 respectively. Of the USA's 72 studies, 49 were focused on HCSO (68.1%), while 21 were focused on DPM (29.2%). The UK, meanwhile, had 29 of 66 studies (43.9%) focused on HCSO, and 35 focused on DPM (53.0%). It is interesting to note that the US and UK place an almost-inverse emphasis on their use of DES; the US prioritized operations over disease management, while the UK focused on disease management over operations. The relative lack of published articles from China is somewhat surprising, with only 14 studies released.

There were 24 countries with only one or two studies published. This indicates that discrete event simulation as a healthcare improvement tool has worldwide application, and is not limited only to developed nations. It also indicates a potential growth opportunity. Developing nations have more flexibility in designing or redesigning their healthcare delivery systems. As such, they have potential for future study and application of DES.

It should be noted that the countries listed are the countries in which the health system of interest is located, not necessarily where the researchers were located or the research was published.

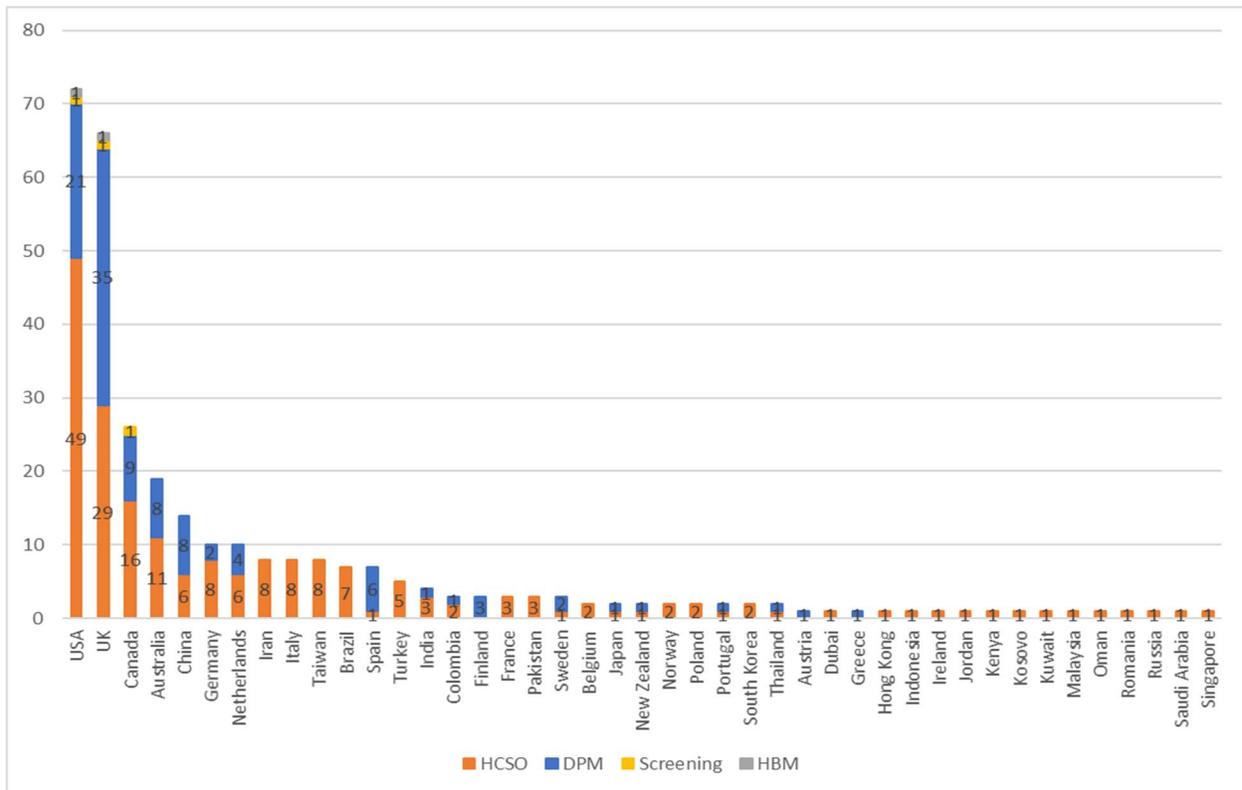

Figure 7. DES application by country of study

### 3.8 Software usage

The initial computerized discrete event simulations in the 1950s were coded directly in FORTRAN. The first dedicated discrete event simulation software package, General Purpose Simulation System (GPSS), was developed by General Electric in 1960. From those early days, all DES software tools have had six common features (Nance, 1996).

   a. Random number generator to represent stochastic uncertainty
   b. Process transformers to convert random numbers to statistical distributions
   c. List processors to add, delete, and manipulate sets and set members
   d. Statistical analysis routines to summarize model behavior
   e. Report generators to present large data outputs
   f. Timing mechanism to explicitly represent the flow of time

From those simple beginnings, there are now over 50 commercially-available discrete event simulation programs. There are even more software tools that can be used in discrete event simulations, though they aren't dedicated DES packages. Dias, Vieira, Pereira and Oliveira (2016) studied the relative popularity of dedicated commercial DES packages. Two limitations to the study was that it only included dedicated DES packages, not multi-purpose tools like Excel and R; and that it considered "popularity," not objective usage rates. "Popularity" was determined as a combination of publication references, database references, social media presence, and web traffic. Their study showed stratification in the popularity of DES tools. Arena was clearly the most popular tool, with ProModel, FlexSim, Simul8, Simio, and AnyLogic in a second tier. Other tools were less popular.

Beyond the somewhat nebulous "popularity" survey, this review looked at which software was used in the published articles (Figure 8). Of the 311 articles screened for review, 47 did not specify the software used in the study. As per the popularity review, Arena was the most popular software. It was also the most-frequently cited software, being used in 49 of the studies. The other more popular software packages were: AnyLogic, ExtendSim, FlexSim, ProModel, Simio, and Simul8. AnyLogic and Simul8 were both in the second tier of most-frequently-cited software, being used 35 and 31 times respectively. Simio, FlexSim, ExtendSim, and ProModel were decidedly third-tier in their published usage, with 11, 11, 5, and 4 articles respectively.

The prior popularity study did not consider non-dedicated DES software. This systematic review indicates a potential area of improvement for future such studies, as R and Excel are quite frequently used, with 32 and 31 citations respectively. Python and Matlab also saw moderate usage, with 14 and 9 citations respectively. Notably, all four of those tools lack the graphical flow visualization of the dedicated DES modeling tools.

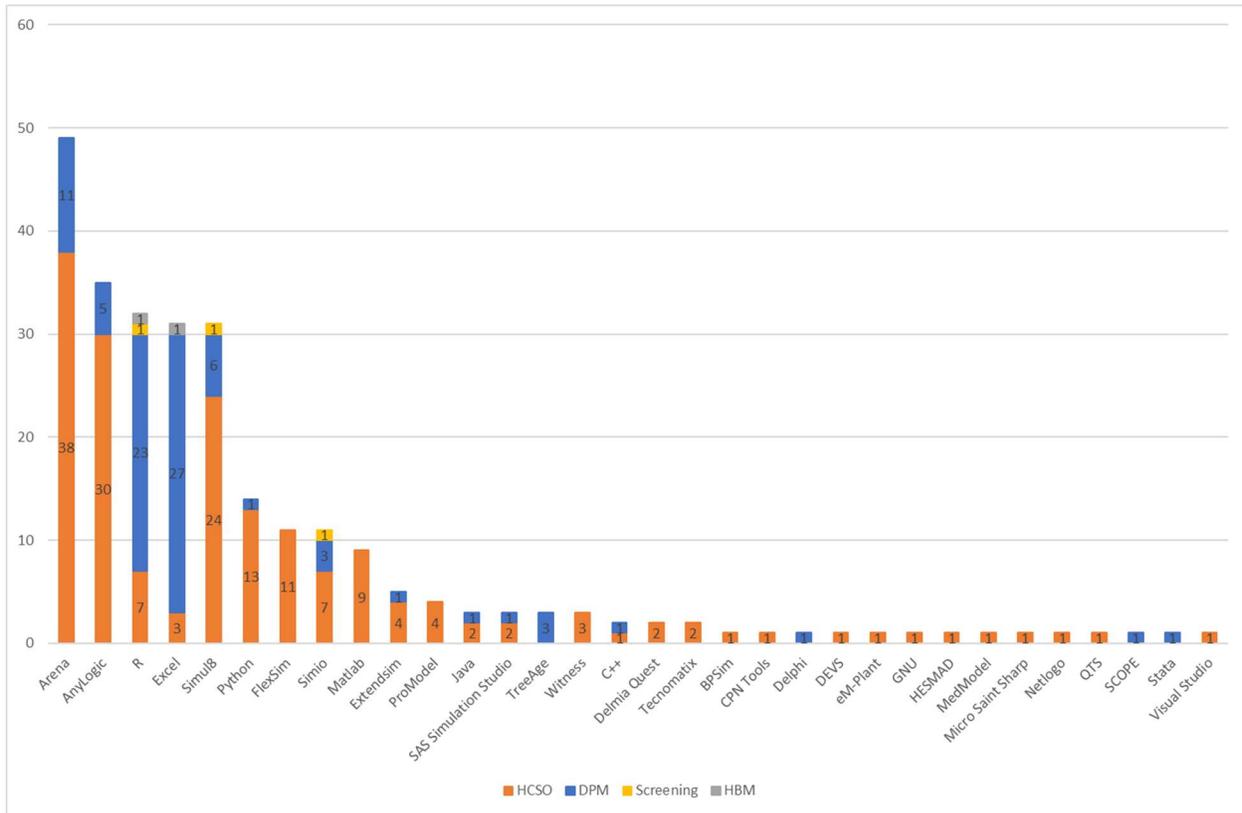

Figure 8. Distribution of DES software usage

It is also instructive to look at the distribution of software used by type of study. For HCSO studies (Figure 9), the dedicated DES tools dominate in cited usage. Arena, AnyLogic, and Simul8 are clearly the most-used tools, with 38, 30, and 24 citations respectively. Those dedicated tools all feature a graphical interface that visualizes the process flow. For HCSO projects, this is particularly useful, as HCSO projects frequently model flows of people or materials in a clinic or hospital, where spatial relations are important to the study.

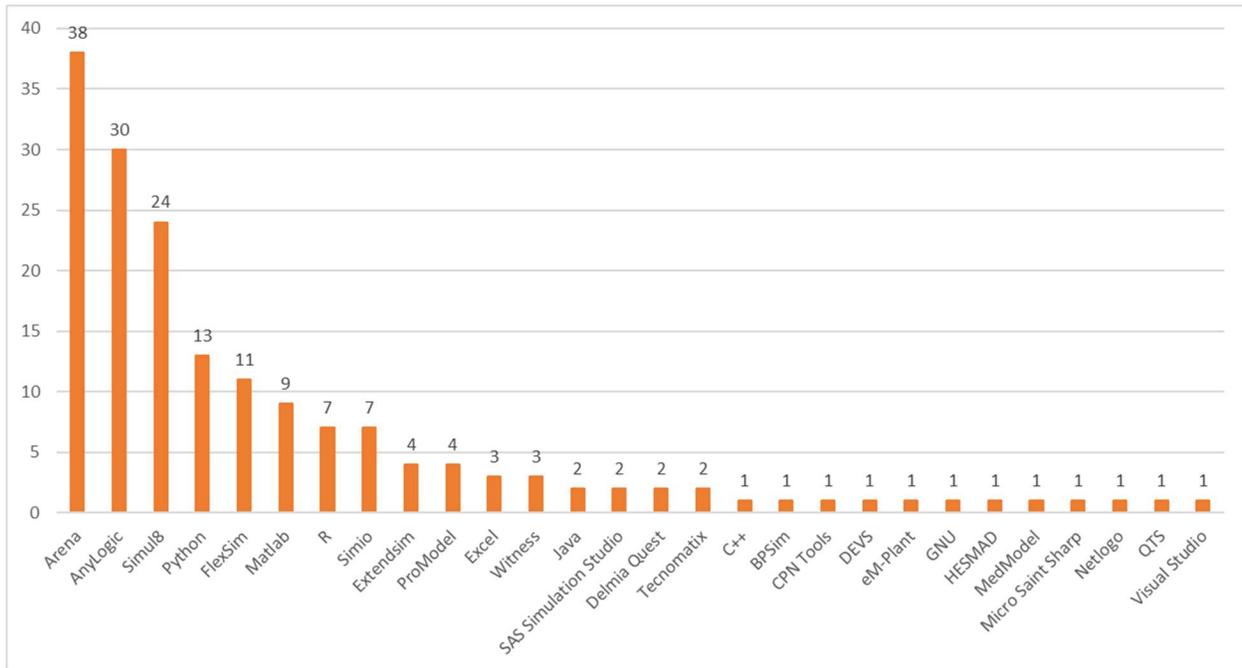

Figure 9. HCSO software usage

As a counterpoint, consider the cited usage of software in DPM projects (Figure 10). In those, R and Excel clearly dominate, with Arena a distant third (with 27, 23, and 11 citations respectively). Simul8, AnyLogic, Simio, and TreeAge make up the next tier, with 6, 5, 3, and 3 citations respectively. In DPM studies, temporal relations are far more important than spatial. In fact, for many DPM studies, spatial relations are not considered at all. When looking at population health outcomes across the entire healthcare system, the ability to visualize entity process flow is much less important than in a HCSO project. As such, graphical DES software loses much of its advantage. The widespread acceptance of R and Excel, combined with their utility for other tasks, makes them good choices for use in DPM studies.

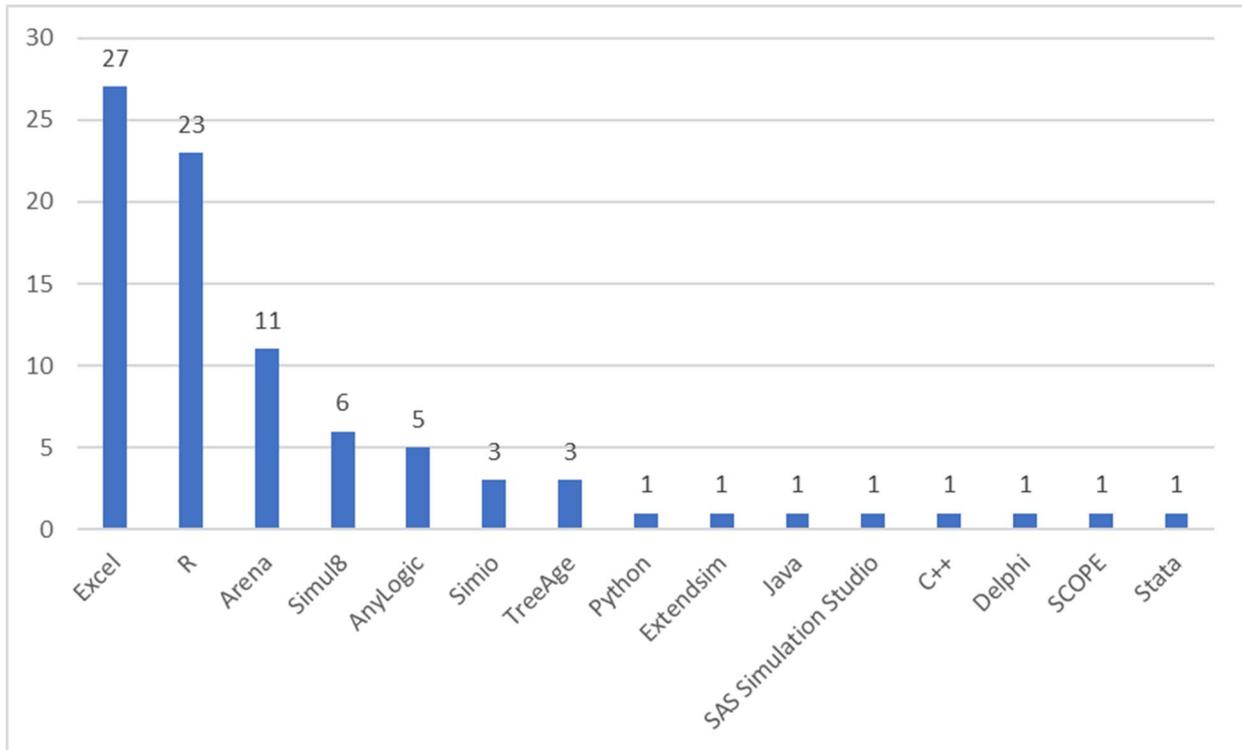

Figure 10. DPM software usage

**3.9 Simulation size**

The size of the system being studied, and the number of entities that pass through it, will affect the speed and computational cost of the discrete event simulation. The size of the health systems simulated varied wildly. See Figure 11. There were operational studies on individual clinics that saw only 308 patients per year (Tofighi *et al*, 2021). That scaled up to a multi-facility regional system that treated 17 million patients per year (Zehrouni *et al*, 2021). The median size of HCSO facilities modeled was 10,000 patients per year.

Similar wide variation was seen in DPM studies. Such studies ranged in size from clinical studies with cohorts as low as 50 patients (Ni & Jiang, 2017), up to studies of the entire health care system, with millions of possible candidates (Gray *et al*, 2017). The median DPM study was 10,000 patients. The great variation in simulation sizes shows the utility of DES in addressing a wide array of scenarios. This variation may be a consequence of the natures of the studies. Typical HCSO models deal with a small-scale operational question, ranging in size from a clinic/department, up to an entire hospital. Meanwhile, DPM models look at entire patient populations across years, thus seeing far more entities pass through the system model.

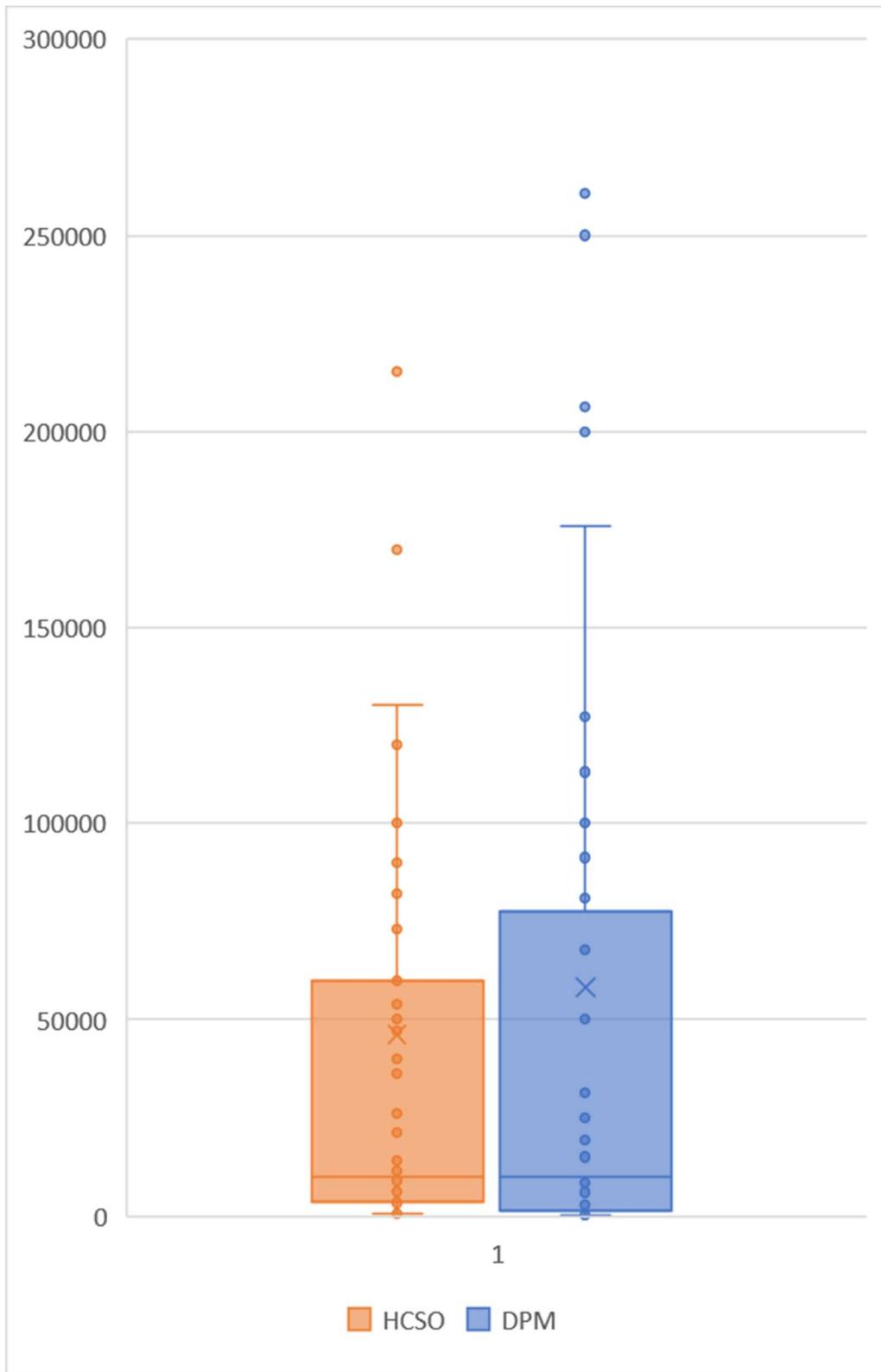

Figure 11. Distribution of health system sizes

**3.10 Patient demographics**

There are variations in practice between adult medicine and pediatric medicine. There is also great variation in patient size in pediatrics, which in turn creates variation in medication dosing, treatment pathways, and process timings. Thus, simulations of processes involving pediatrics would often be expected to show higher variation than in the adult population. The vast majority of studies were focused on adult patients; 298 of 311 (95.8%) studies were adult. This is not surprising for DPM models. See table 2. As shown later in section 4.2.2, the majority of disease processes studied traditionally affect adults, and especially geriatrics, more so than pediatrics. Of the HCSO studies around pediatrics, the primary areas of focus were patient scheduling and resource allocation. There was one study specifically focused on family medicine (that is, the treatment of both children and adults), looking at capacity planning.

|           | Age   |           |        |
|-----------|-------|-----------|--------|
| **Focus** | **Adult** | **Pediatric** | **Family** |
| **HCSO**  | 190   | 9         | 1      |
| **DPM**   | 103   | 3         |        |
| **Screening** | 3 |           |        |
| **HBM**   | 2     |           |        |

Table 2. Distribution of studies by target age

The bulk of models focus on adult patients exclusively. This is partly due to opportunity - there are more and more-varied adult healthcare providers. This is to be expected, considering how people access healthcare over the course of their lives. Barring an early chronic condition, young people don't seek healthcare frequently; conversely, even nominally healthy adults will require additional healthcare resources as they age, particularly those associated with cancer, cardiovascular health, and gerontology. Another potential factor is the variation in pediatric patients, who range from birth to age 18. The wide variation in size and metabolic function makes it challenging to code simulation models, with additional attributes that must be tracked and programmed to account for the difference in patient populations. However, those discrepancies between pediatric and adult patients also indicate an area of potential opportunity of expansion and further study.

Of the nine pediatric HCSO simulations, four were focused on emergency department operational changes. This type of model is not terribly different from an adult-only model. Two of them studied the effect of patient scheduling - one in a clinic setting, the other in a surgical setting. One model studied the effect of staff scheduling, specifically in a neonatology clinic. The other two investigated operational change and resource allocation in a pediatric clinic.

## 4. Discussion

The search for this review used two databases, PubMed and EBSCOhost. Additional databases exist covering healthcare economics such as Embase, NHSEED and HTA, while Medline is a subset of PubMed (www.nlm.nih.gov/bsd/difference.html). However, the databases we used are comprehensive and for both to have missed much of significance seems unlikely, and thus should be sufficient to support our findings and conclusions.

The selection criteria could also be a limiting factor in gathering articles. By only surveying peer-reviewed journal articles, some percentage of the academic work has undoubtedly been missed. Some of the work in healthcare improvement via discrete event simulation is not being written in academic journals, or in some cases, publicized in any fashion. However, as before, it is reasonable to conclude that the body of literature included contains most of the important work, and is a fair overview of the majority of the work being done.

Finally, this review did not investigate the quality of the outcomes of the discrete event simulations. A good way for an article describing a particular DES study to report on the quality of its findings is the CHEERS or Consolidated Health Economic Evaluation Reporting Standards (Husereau et al., 2022). Similarly CHEERS could, although it is outside the scope of this report, potentially be used in a meta-analysis to compare the usefulness of DES across various categories of DES applications. Even where the only outcome of a study is increased certainty that the current process is superior to a proposed alternative, there should be a defined outcome as a result of simulation.

## 5. Conclusions

Discrete event simulation is a stochastic, dynamic, discretely-changing modeling method. It models the movement of entities through a transformative process over time. Well-defined models, which have clearly defined objectives and are verified and validated with input from process owners and decision makers, are capable of assisting decision makers understand their system, the impact of possible changes, and therefore make informed decisions to improve the system. It is well suited to analyzing a wide variety of scenarios, but the computational requirements can limit the size of the models.

A wide array of software packages are available for discrete event simulation modeling. While many have graphical interfaces, allowing for visualization of flow through the system, that isn't a requirement. Each software has its own nuances, such as grammar and syntax, and requires the designer to learn the specifics of the tool.

The use of discrete event simulation in healthcare has continued to grow. Applications in healthcare system operations are the dominant use of discrete event simulation in the field. This should not come as a surprise, as those types of models are very similar to the types of systems that DES was designed to address.  With those models, resource allocation and process efficiency are the most frequent objectives.  While not originally designed to model clinical outcomes, the continued expansion of discrete event simulation to clinical simulations shows its adaptability and flexibility, along with its utility.  Most clinically-oriented studies look at the long-term economic and quality-of-life outcomes for various treatments, such as cancer and cardiovascular disease.

While much work has been done at an operational level and at a clinical level, there is little evidence of simulation applied to healthcare at a strategic level.  There is much focus on operational flow and efficiency through individual facilities, and analyses of which treatment pathways generate the best long-term health outcomes, but little on modeling which system-level healthcare policies will affect population-level access to healthcare, or how those policies will affect the quality of care.  Given the turbulent nature of national-level healthcare policy, especially in the United States, this is unfortunate.

Patient and staff scheduling have both been studied, but the intersection between the two has not.  Typically, either the patient population or the staffing pattern remains fixed, and the other changed to find an ideal balance.  A more robust approach would be to identify an objective function taking both into account, and using DES to find the optimal balance of both.

## 6. List of abbreviations

ABM: Agent-Based Model
DES: Discrete Event Simulation
DPM: Disease Progression Management
ED: Emergency Department
EMT: Emergency Medical Transport
GPSS: General Purpose Simulation System
HBM: Health Behavior Modeling
HCSO: Health Care System Operations
ISPOR-SMDM: The Professional Society for Health Economics and Outcomes Research (formerly International Society for Pharmacoeconomics and Outcomes Research) Society for Medical Decision Making
PDSA: Plan-Do-Study-Act
PRISMA: Preferred Reporting Items for Systematic Reviews and MetaAnalyses
QALY: Quality-Adjusted Life Years

## 7. Declarations

Ethics approval and consent to participate
Not applicable

Consent for publication
Not applicable

Availability of data and materials
The datasets used and/or analyzed during the current study are available from the corresponding author on reasonable request.

Competing interests
The authors declare that they have no competing interests.

Funding
The author received no financial support for the research, authorship, and/or publication of this article.

Authors' contributions
The sole author was responsible for all aspects of the study and manuscript. The author read and approved the final manuscript.

Acknowledgements
The author is grateful to Daniel Berleant (DB) for feedback on the draft and suggestions provided in the reviewing process.

Authors' information
John J Forbus
BSME University of Arkansas at Fayetteville
MSSE University of Arkansas at Little Rock
Process Improvement Engineer, Arkansas Children's Hospital, Little Rock Arkansas

Criddle, J; Holt J E. "Use of Simulation Software in Optimizing PACU Operations and Promoting Evidence-Based Practice Guidelines." *Journal of PeriAnesthesia Nursing*, 2018, vol.33, no.4, pp.420-425. doi:10.1016/j.jopan.2017.03.004

Dall, T M; Gallo, P D; Chakrabarti, R; West, T; Semilla, A P; Storm, M V. "An Aging Population And Growing Disease Burden Will Require ALarge And Specialized Health Care Workforce By 2025." *Health Affairs*, 1 Nov 2013, www.healthaffairs.org/doi/abs/10.1377/hlthaff.2013.0714

Dias, L; Vieira, A; Pereira, G; Oliveira, J. "Discrete Simulation Software Ranking – A Top List of the Worldwide Most Popular and Used Tools." *Proceedings of the 2016 Winter Simulation Conference*, 2016

Dieleman, J M; Myles, P S; Bulfone, L; Younie, S; van Zaane, B; McGiffin, D; Moodie, M; Gao, L. "Cost-effectiveness of Routine Transoesophageal Echocardiography During Cardiac Surgery: A Discrete-Event Simulation Study." *British Journal of Anaesthesia*, 2020, vol.124 no.2, p.136-145.

Donnelly, P; Kirk, P. "Use the PDSA Model for Effective Change Management." *Education for Primary Care*, 2015, vol.26, no.4, pp. 279-281. doi:10.1080/14739879.2015.11494356

Dutta, D; Parry, F; Obaid, M; Ramadurai, G. "Mechanical Thrombectomy in Stroke - Planning For Service Expansion Using Discrete Event Simulation." *Future Healthcare Journal*, 2020, vol.7, no.1, p.65-71

Elizandro, D; Matson, J. "Discrete Event Simulation Using Excel/VBA." *Proceedings of the 2005 American Society for Engineering Education Annual Conference & Exposition*, 2005

Furushima, D; Yamada, H; Kido, M; Ohno, Y. "The Impact of One-Dose Package of Medicines on Patient Waiting Time in Dispensing Pharmacy: Application of a Discrete Event Simulation Model." *Biological and Pharmaceutical Bulletin*, 2018, vol.41, no.3, pp.409-418

Gray, E; Donten, A; Karssemeijer, N; van Gils, C; Evans, D G; Astley, S; Payne, K. "Evaluation of a Stratified National Breast Screening Program in the United Kingdom: An Early Model-Based Cost-Effectiveness Analysis." *Value in Health*, 2017, vol.20, no.8, pp.1100-1109 de Koning, H; Verver, J P; van den Heuvel, J; Bisgaard, S; Does, R J. "Lean Six Sigma in Healthcare." *Journal for Healthcare Qualit*y, 2006, vol.28, no.2, pp. 4-11. doi:10.1111/j.1945-1474.2006.tb00596.x
27

Nau, C; Kumanyika, S; Gittelsohn, J; Adam, A; Wong, M S; Mui, Y; Lee, B Y. "Identifying Financially Sustainable Pricing Interventions to Promote Healthier Beverage Purchases in Small Neighborhood Stores." *Preventing Chronic Disease,* 2018, vol.15

Nawaz, R; Maqsood, S; Baber, A R. "Analysis of Emergency Medical Response Service in Peshawar through Simulation." *Mehran University Research Journal of Engineering and Technology*, 2019, vol.38, no.4, pp.1033-1044

Ni, W; Jiang, Y. "Evaluation on the Cost-Effective Threshold of Osteoporosis Treatment on Elderly Women in China Using Discrete Event Simulation Model." *Osteoporosis International*, 2016, vol.28, no.2, pp.529-538

Nguyen, L K N; Megiddo, I; Howick, S. "Simulation Models For Transmission of Health Care-Associated Infection: A Systematic Review." *American Journal of Infection Control*, 2020, vol.48, no.7, pp.810-821. doi:10.1016/j.ajic.2019.11.005

Pennington, B; Filby, A; Owen, L; Taylor, M. "Smoking Cessation: A Comparison of Two Model Structures." *PharmacoEconomics,* 2018, vol.36, no.9, pp.1101-1112.

Robinson, S. "Discrete-Event Simulation: from the Pioneers to the Present, What next?" *Journal of the Operational Research Society*, 2005, vol.56, no.6, pp.619-629. doi:10.1057/palgrave.jors.2601864

Schroer, B J; Tseng, F T. "Modelling Complex Manufacturing Systems Using Discrete Event Simulation." *Computers & Industrial Engineering*, 1988, vol.14, no.4, pp.455-464. doi:10.1016/0360-8352(88)90047-2.

Sewell, B; Jones, M; Gray, H; Wilkes, H; Lloyd-Bennett, C; Beddow, K; Bevan, M; Fitzsimmons, D. "Rapid Cancer Diagnosis For Patients With Vague Symptoms: A Cost-effectiveness Study." *The British Journal of General Practice : The Journal of the Royal College of General Practitioners*, 2020; vol.70(692) e186-e192. doi:10.3399/bjgp20X708077

Sharma, P. "Discrete-Event Simulation." *International Journal Of Scientific & Technology Research*, 2015, vol.4, no.04.

Shi, Y; Graves, J A; Garbett, S P; Z, Z; M, R; Wang, X; Harrell, F E; Lasko, T A; Denny, J C; Roden, D M; Peterson, J F; Schildcrout, J S. "A Decision-Theoretic Approach to Panel-Based, Preemptive Genotyping." *MDM Policy & Practice,* 2019-08, vol.4, no.2. 2381468319864337–2381468319864337.
29

3131